%% file: fmcad_submission.tex
\pdfoutput=1
\documentclass[11pt, conference, compsocconf]{IEEEtran}
\ifCLASSINFOpdf
\else
\fi
\usepackage[dvips]{graphicx}
\usepackage{latexsym}
\usepackage{epsf}
\usepackage{epsfig}

\usepackage{amsmath} 
\usepackage{amssymb}
\usepackage{amsthm}  
\usepackage{amsxtra}
\usepackage{amsxtra}
\usepackage{amsfonts}

\usepackage{enumerate}
\usepackage{amsthm}
\usepackage{amsbsy}

\usepackage{tikz}

\usepackage{algorithm}
\usepackage{algpseudocode}

\usepackage{graphics}
\usepackage{color}
\usepackage{graphicx}
\usepackage{enumerate}
\usepackage{fancyvrb}
\usepackage{alltt}
\usepackage{url}

\usepackage{colortbl}
\usepackage{multirow}
\usepackage{chngpage}

\usepackage{amsfonts}
\usepackage{pifont}

\usepackage{relsize}

\include{macros}

\begin{document}
%
\title{Synthesis of Sequential Extended Regular Expressions for Verification}

\author{\IEEEauthorblockN{Mohamad Noureddine\IEEEauthorrefmark{1} \hspace*{2em} Fadi A. Zaraket \IEEEauthorrefmark{1} \hspace*{2em} Ali S. Elzein\IEEEauthorrefmark{2} }
  \IEEEauthorblockA{\IEEEauthorrefmark{1} American University of Beirut,\{man17,fz11\}@aub.edu.lb}
  \IEEEauthorblockA{\IEEEauthorrefmark{2} IBM Systems and Technology, elzein@us.ibm.com }
}


%


\maketitle

\begin{abstract}
 Synthesis techniques take realizable Linear Temporal Logic specifications 
 and produce correct circuits that implement the specifications. 
 The generated circuits can be used directly, or 
 as miters that check the correctness of a logic design. 
 Typically, those techniques generate non-deterministic finite state
 automata, which can be determinized at a possibly exponential cost. 
 Recent results show multiple advantages of using deterministic 
 automata in symbolic and bounded model checking of LTL safety properties. 
 In this paper, we present a technique with a supporting tool
 that takes a {\em sequential extended regular expression} specification $\Phi$, 
 and a logic design implementation $S$, 
 and generates a {\em sequential circuit} $C$, 
 expressed as an {\em And-Inverted-Graph},
 that checks whether $S$ satisfies $\Phi$. 
 The technique passes the generated circuit $C$ to 
 ABC, a bounded model checker, to validate correctness.

 We use {\em free input variables } to encode the non-determinism in $\Phi$ 
 and we obtain a number of states in miter linear in the size of $\Phi$. 
 Our technique succeeds to generate the input to the model checker 
 while other techniques fail because of the exponential blowup,
 and in most cases, ABC succeeds to either find defects in the design
 that was otherwise uncheckable, or validate the design. 
 We evaluated our technique against several industrial benchmarks including 
 the IBM arbiter, a load balancer, and a traffic light system, and
 compared our results with the NuSMV framework. 
 Our method found defects and validated systems NuSMV could not 
 validate.

\end{abstract}


%
\IEEEpeerreviewmaketitle

\section{Introduction} \label{s:intro}
\input{introduction}

\section{Preliminaries} \label{s:prel}
\input{background}

\section{Motivating example} \label{s:motivation}
\input{motivation}

\section{Equisatisfiable DFA } \label{s:core}
\input{construction}

\section{Implementation} \label{s:implementation}
\input{implementation}

\section{Experimental Results} \label{s:results}
\input{results}

\section{Related work} \label{s:related}
\input{related}

\section{Conclusion} \label{s:conclusion}
\input{conclusion}



\bibliographystyle{IEEEtran}
\bibliography{fmcad_submission}
%
%


\end{document}

%% file: macros.tex
\usepackage{xspace}

\newcommand{\MATH}[1]{\ensuremath{#1}\xspace}
\newcommand{\md}[1]{\MATH{[ #1 ]}}                               

\newcommand{\remove}[1]{}

\newcommand{\cci}[1]{{\small\texttt{#1}}}

\newcommand{\ap}{\ensuremath{\mathbb{A}}}
\newcommand{\eval}{\ensuremath{\mathbb{V}}}

\newtheorem*{definition}{Definition}
\newtheorem{theorem}{Theorem}

\renewcommand{\le}{\leqslant}

\newcommand{\set}[1]{\left\{ #1 \right\}}

\newcommand{\tpl}[1]{\ensuremath{\langle #1 \rangle}}

\newcommand{\union}{\bigcup}

\newcommand{\terms}{\ensuremath{{\mathbb T}}\xspace}

\newcommand{\ang}[1]{\ifmmode{\left\langle #1 \right\rangle}
   \else{$\left\langle${#1}$\right\rangle$}\fi}

\newcounter{lctr}

\newcommand{\pseudocode}[1]{\ensuremath{\mathbf{#1}\ }}

\renewcommand{\int}{\pseudocode{int}}

\newcommand{\true}{\ensuremath{\mathit{true}}}
\newcommand{\false}{\ensuremath{\mathit{false}}}

\newcommand{\beqn}{\begin{centeqn}}
\newcommand{\eeqn}{\end{centeqn}}

\newcommand{\bleqn}[1]{\begin{centlabeqn}{#1}}
\newcommand{\eleqn}{\end{centlabeqn}}

\newenvironment{centeqn}  
   {{\ss\\ \hspace*{\fill}}}
   {\hspace*{\fill}\ss\\}

\newsavebox{\EqnLabel}
\newenvironment{centlabeqn}[1]  
   {\sbox{\EqnLabel}{#1}
     {\medskip\\ \hspace*{\fill}}
   }
   {\hfill{\makebox[0in][r]{\usebox{\EqnLabel}}}\medskip\\}

\newenvironment{centeqn-nbsp} 
   {{\medskip\\ \hspace*{\fill}}}
   {\hspace*{\fill}}

%% file: introduction.tex

Safety critical systems such as medical and navigation control devices 
rely on digital systems in order to provide accurate services. 
Verification techniques, such as symbolic and bounded model checking,
address the correctness of digital systems with respect
to formal specifications written in languages such 
as {\em linear temporal logic } (LTL).
{\em Sequential extended regular expressions} (SERE) form a subset of the 
{\em Property Specification Language} (PSL) that 
constitute a practical way to specify
logic designs~\cite{psl2004}. 
SERE covers a practical subset of LTL. 

Automated synthesis tools such as Wring~\cite{p:wring},
Lily~\cite{jobstmann2006optimizations},
and UNBEAST~\cite{ehlers2010symbolic} 
take an LTL specification and generate a correct 
implementation. 
Validation tools such as
Focs~\cite{p:Focs}, 
NuSMV~\cite{cimatti2002nusmv}, 
and SPIN~\cite{holzmann1997model}
take a specification and an implementation 
and check whether the implementation satisfies the 
specification. 
They either provide a proof of correctness, 
a counterexample, or 
an inconclusive result when they reach their 
computational boundaries~\cite{kropf1999introduction}.

NuSMV~\cite{cimatti2002nusmv} and COSPAN~\cite{hardin1996cospan}
typically translate the design $S$ and the
negation of the LTL specification $\Phi$ into non-deterministic finite
state automata (NDFA) $M_S$ and $M_{\lnot{}\Phi}$
(typically using B\"uchi automata), respectively, and 
then perform symbolic model checking
on the resulting cross product automaton~\cite{vardi2007automata, clarke1999model}. 
This results in an online determinization of the 
assertion automaton and thus the 
state space explosion problem is inherent to symbolic model checking~\cite{vardi2007automata}. 

The majority of the LTL properties to be verified are safety properties, 
to which finite violating counterexamples can be found. 
Therefore, researchers consider translating the LTL specifications into deterministic finite 
state automaton (DFA)~\cite{vardi2007automata, armoni2005efficient} risking 
the state space explosion problem~\cite{sipser2006introduction}.
This often limits the ability of synthesis tools to generate input to 
the model checkers for verification. 
NuSMV uses several abstraction and reduction techniques, such as the 
cone of influence reduction~\cite{berezin1998compositional} and 
other Binary Decision Diagrams (BDD) based techniques~\cite{ranjan1995efficient}, 
in order to avoid such a problem. 

In this paper, we present a technique and a supporting 
tool that takes 
an SERE specification $\Phi$ and an implementation of it $S$,
and generates a sequential circuit
$C$ that checks whether $S$ satisfies $\Phi$.
Our technique encodes the non-determinism in $\Phi$ 
using additional free 
variables, and generates an {\em equisatisfiable} 
sequential circuit $C_{\Phi}$ 
such that $C_{\Phi}$ has a number of
states linear in the size of $\Phi$. 
Informally, a sequential circuit $C$ with 
a designated output $o$ therein
is equisatisfiable to an SERE specification $\Phi$ 
when $o$ is satisfiable if and only if $\Phi$ is satisfiable.
The circuit $C_{\Phi}$ can not be used as an 
implementation of $S$
and can only be used as a miter in model checking tools
to validate an implementation of $\Phi$. 
The technique translates the implementation 
$S$ into a sequential circuit $C_S$
and builds $C$ as the composition of 
$C_{\Phi}$ and $C_S$. 
The technique then applies the ABC model checker on 
the generated
sequential circuit $C$ and checks for correctness. 

Encoding non-determinism using free variables is a textbook technique~\cite{sipser2006introduction}.
Reportedly, it might have been used in existing tools such as ``smvtoaig'' for a limited 
subset of the ``smv'' designs. 
Up to our knowledge, we are the first to use this technique in an open source tool 
to enable the verification of logic design against SERE specifications.
Our technique enables the ABC model checker to find defects and prove the correctness 
of systems where it is not possible with existing techniques. 

We implemented and evaluated our techniques with benchmarks 
from UNBEAST~\cite{ehlers2010symbolic} and LILY~\cite{jobstmann2006optimizations},
in addition to the IBM arbiter presented in~\cite{IBM_arbiter}.
We provide our tool, the appendices of the paper including proofs, 
and the benchmarks for the experiments online
~\footnote{\label{fn:url}\url{http://webfea.fea.aub.edu.lb/fadi/dkwk/doku.php?id=ltlsyn}}.
Our technique was able to find problems in several designs where it was not possible before.
The supporting tool allows the user to
\begin{itemize}
  \item {\em prove} that an implementation
  satisfies an SERE property using 
  satisfiability and bounded model checking,
\item {\em debug } the implementation and the 
    specification using the generated counterexample, and 
\item {\em simulate} the implementation 
  and the specification and inspect
  the results. 
\end{itemize}

The rest of this paper is organized as follows. Section~\ref{s:prel} presents some preliminary information,
Section~\ref{s:motivation} motivates our approach using a simple example. The core of the synthesis technique
is presented in section~\ref{s:core}. We describe our implementation in Section~\ref{s:implementation}, and 
show a summary of the experimental results in Section~\ref{s:results}.
Related work is summarized in Section~\ref{s:related} and we conclude in Section~\ref{s:conclusion}.

%% file: background.tex
Let \ap~ be a set of atomic propositions. 
The mapping 
$\ap \to \mathbb{B}$ denotes
a valuation to the atomic propositions 
in \ap\
where $\mathbb{B}=\set{\true, \false}$. 
Let $\eval= (\ap \rightarrow \mathbb{B})$ be the set 
of all such valuations. 

SERE formulae range over the alphabet 
$\Sigma=\ap\ \union~\{;,*, \land,\lor,\lnot,(,)\}$,
where (1) $\land, \lor$ are Boolean binary operators
denoting conjunction and disjunction, respectively, 
(2) `;' is a sequential binary operator denoting temporal next, 
(3) $\lnot$ is a Boolean unary operator denoting logical negation, and
(4) $*$ is a sequential unary operator denoting zero or more times.

\begin{definition}[SERE terms]
\label{def:sere-term}
An atomic proposition in \ap is an {\em SERE term}. 
If $t_1$ and $t_2$ are SERE terms, then 
$t_1 \land t_2$, $t_1 \lor t_2$, $(t_1)$ and $\lnot t_1$ are 
SERE terms. 
We denote by \terms the set of all SERE terms. 
\end{definition}


\begin{definition}[SERE formula]
\label{def:sere-term}
An SERE term is an SERE formula. 
Given $\phi$ and $\psi$ are SERE formulae and $t$ is
an SERE term, then $t*$,
$\phi ; \psi$, $(\phi)$ , 
$\phi \land \psi$, and $\phi \lor \psi$ are
all SERE formulae. 
We denote by SERE the set of all SERE formulae. 
\end{definition}


A valuation $v \in \eval$ satisfies an atomic proposition $a \in \ap$, 
$v \models a$ iff $v$ maps $a$ to \true; we denote that also by 
$v(a)=\true$. 
A trace 
$\rho=\tpl{v_1, v_2, \ldots, v_i, \ldots, v_n}, i \in [1 \ldots n]$ is a sequence of valuations.
We denote by (1) $\rho=\rho_1\circ\rho_2$ the concatenation
of the traces $\rho_1$ and $\rho_2$, and (2)
$\rho(t,i)$ the value of term $t$ at the $i^{th}$ entry of $\rho$. 

\begin{definition}[SERE term semantics]
\label{def:sere-term-sem}
Let $\rho$ be a trace, and let $e_1$ and $e_2$ be 
SERE terms. 
\begin{itemize}
  \item If $e_1 \in \ap$ then $\rho \models e_1$ iff 
    $|\rho| = 1$ and $\rho(e_1,1)$
  \item $\rho \models \lnot e_1$ iff $|\rho| = 1$ and $\rho \not\models e_1$
  \item $\rho \models e_1 \land e_2$ iff $\rho \models e_1$ and $\rho \models e_2$ 
  \item $\rho \models e_1 \lor e_2$ iff $\rho \models e_1$ or $\rho \models e_2$
\end{itemize}
We denote by \md{e} all the valuations that 
satisfy term $e$. 
\end{definition}

Let $\psi$ be an SERE formula of the form  
$x_1y_1; x_2y_2; \ldots; x_ny_n$ where $i\in [1 \ldots n]$
$x_i \in \terms$, $y_i \in \set{\epsilon, *}$,
and $\epsilon$ is the empty string. 


\begin{definition}[SERE formula semantics]
\label{def:sere-for-sem}
Let $\rho$ be a trace and $\psi$ be an SERE formula. 
We say $\rho$ satisfies $\psi (\rho \models \psi)$ in the following cases. 
\begin{itemize}
  \item $\psi = x_1y_1, y_1 = \epsilon$ iff $|\rho| = 1$ and $\rho(x_1,1)$
  \item $\psi = x_1y_1, y_1 = *$ iff 
    \begin{itemize}
      \item $\rho = \epsilon$, 
      \item $|\rho|=1$ and $\rho \models x_{1}$, or 
      \item there exists traces $\rho_1,\rho_2$ such that 
           $\rho_1 \neq \epsilon$, $\rho = \rho_1\circ\rho_2$, 
           $\rho_1 \models \psi$, and $\rho_2 \models \psi$
    \end{itemize}
    \item $\psi = x_1y_1;x_2y_2$ iff there exists traces 
      $\rho_1,\rho_2$ such that 
      $\rho=\rho_1\circ\rho_2$, $\rho_1 \models x_1y_1$, and
      $\rho_2 \models x_2y_2$
    \item $\psi=\phi_1;\phi_2$ where $\phi_1,\phi_2$ are formulae iff
      there exists traces $\rho_{\mathit{prefix}},\rho_1,\rho_2,\rho_{\mathit{suffix}}$,
      such that 
      $\rho=\rho_{\mathit{prefix}}\circ\rho_1\circ\rho_2\circ\rho_{\mathit{suffix}}$,
      $\rho_1 \models \phi_1$,
      and $\rho_2 \models \phi_2$
    \item $\psi = \phi_1 \land \phi_2$ iff $\rho \models \phi_1$ and $\rho \models \phi_2$
    \item $\psi = \phi_1 \lor \phi_2$ iff $\rho \models \phi_1$ or $\rho \models \phi_2$
\end{itemize}
We denote by \md{\psi} all the traces that satisfy formula $\psi$. 
\end{definition}

\begin{definition}[Deterministic finite state automata]
\label{def:dfsa}
A deterministic finite state automata (DFA) is a tuple 
$M=(Q, I, F, \Sigma, L, \delta)$ where $Q = \{s_0, s_1, \ldots, s_n\}$ is 
the set of states of $M$, $I \subseteq Q$ is the set of initial states, 
$F \subseteq Q$ is the set of accept states, 
$\Sigma=\eval$ is the input alphabet of $M$,
$L\subseteq \terms$ is a set of transition labels 
such that $\delta = Q \times L \rightarrow Q$ is 
the state transition function.
Note that labels with joint alphabet symbols
(e.g. $a$, $a \land b$, $a \lor b$) are not allowed
on edges outgoing from a state $s$ in order to keep the
transitions deterministic. 
\end{definition}

%
%
The semantics of DFA are defined in the typical manner.
A sequence of input valuations $\rho=\tpl{v_0,v_1,\ldots,v_{n-1}}$, 
determines a sequence of state transitions $\sigma=\tpl{s_0,s_1,\ldots,s_n}, s_0\in I$,
and $s_{i+1} = \delta (s_i, e)$ 
where $e\in L$ and $v_i \in \md{e}$. 
We say $\rho$ satisfies $M$ ($\rho \models M$) iff $s_n$ is 
an accept state of $M$; ($s\in F$).



\begin{definition}[Equisatisfiability]
  \label{def:equisat}
 We say a DFA $M$ is equisatisfiable to an SERE formula $\psi$ iff 
  $M$ is satisfiable iff $\psi$ is satisfiable. 
  That is $\exists \rho. \rho \models M \Leftrightarrow \exists \rho'. \rho' \models \psi$.
\end{definition}


%% file: motivation.tex

\begin{figure}[bt]
\centering
\resizebox{.75\columnwidth}{!}{
    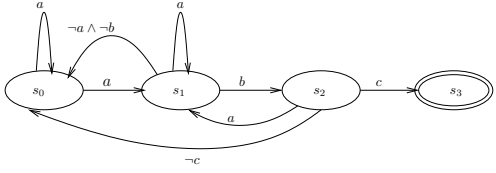
}
\caption{NDFA of a;b;c}
\label{f:mot1_ndfa}
\end{figure}

Consider the SERE formula $\psi = a;b;c$.
NDFA $M$ in Figure~\ref{f:mot1_ndfa} simulates $\psi$ with non-deterministic
transitions in its initial state. 
Once $M$ receives a valuation where $a$ is \true, 
it can move into state $s_1$ or remain in $s_0$ since
$\delta{}(s_0, a) = \set{s_0, s_1}$.

\begin{figure}[t]
 \centering
 \resizebox{0.8\columnwidth}{.15\textheight}{
     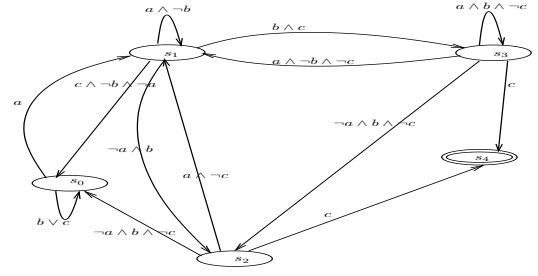
 }
 \caption{DFA of a;b;c using JFLAP~\cite{rodger2006jflap}}
 \label{f:mot1_dfa1}
\end{figure}

Typically, an NDFA $M$ is translated into a DFA $M'$ using 
{\em subset construction} with a possible exponential
blowout in the number of states~\cite{sipser2006introduction}.  
In brief, states in $M'$ are subsets of states in $M$ 
and transitions are constructed to make $M'$ equivalent to $M$,
yet deterministic. 
Figure~\ref{f:mot1_dfa1} shows a DFA equivalent to
$M$ produced using the JFLAP tool~\cite{rodger2006jflap}. 

\begin{figure}[bt]
\centering
\resizebox{.75\columnwidth}{!}{
    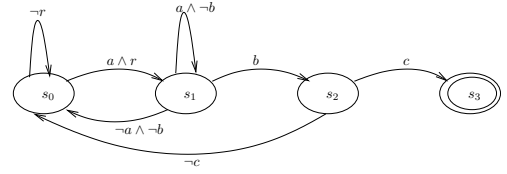
}
\caption{DFA of a;b;c with free atomic proposition r}
\label{f:mot1_dfa}
\end{figure}

Instead, 
we encode the non-determinism using an additional 
free atomic proposition $r$ as shown in 
in Figure~\ref{f:mot1_dfa}. 
This results in DFA $M^a$ that is equisatisfiable to $psi$ 
and that has a number of states linear in the 
number of terms in $\psi$.
We use $M^a$ with symbolic and 
bounded model checkers wherever it is expensive or 
impossible to generate an equivalent DFA for $\psi$.
Our technique 
leaves it to the model checker
to efficiently handle the free variables added by our 
synthesis technique.
In practice, even though our technique does not reduce the 
inherent complexity of the problem,
it enables the application of several reduction and abstraction 
transformations available in model checkers such as ABC 
to reduce and solve the problem. 
These are not applicable without our technique. 



Notice that, for each trace $\rho$ that satisfies $a;b;c$,
there is a trace of $r$ values that makes $\rho$ satisfiable for 
the DFA in Figure~\ref{f:mot1_dfa}. 
In particular, set $r$ to \true\ where the matching sequence 
starts in $\rho$ and to \false\ otherwise. 

%% file: figures/ex1_ndfsm.pdftex_t
\begin{picture}(0,0)%
\includegraphics{figures/ex1_ndfsm.pdf}%
\end{picture}%
\setlength{\unitlength}{4144sp}%
\begingroup\makeatletter\ifx\SetFigFont\undefined%
\gdef\SetFigFont#1#2#3#4#5{%
  \reset@font\fontsize{#1}{#2pt}%
  \fontfamily{#3}\fontseries{#4}\fontshape{#5}%
  \selectfont}%
\fi\endgroup%
\begin{picture}(5641,1978)(1793,-3896)
\put(2926,-2941){\makebox(0,0)[lb]{\smash{{\SetFigFont{12}{14.4}{\familydefault}{\mddefault}{\updefault}{\color[rgb]{0,0,0}$a$}%
}}}}
\put(3781,-2041){\makebox(0,0)[lb]{\smash{{\SetFigFont{10}{12.0}{\familydefault}{\mddefault}{\updefault}{\color[rgb]{0,0,0}$a$}%
}}}}
\put(5356,-3031){\makebox(0,0)[lb]{\smash{{\SetFigFont{10}{12.0}{\familydefault}{\mddefault}{\updefault}{\color[rgb]{0,0,0}$s_2$}%
}}}}
\put(6076,-2941){\makebox(0,0)[lb]{\smash{{\SetFigFont{10}{12.0}{\familydefault}{\mddefault}{\updefault}{\color[rgb]{0,0,0}$c$}%
}}}}
\put(6931,-3031){\makebox(0,0)[lb]{\smash{{\SetFigFont{10}{12.0}{\familydefault}{\mddefault}{\updefault}{\color[rgb]{0,0,0}$s_3$}%
}}}}
\put(4366,-3346){\makebox(0,0)[lb]{\smash{{\SetFigFont{10}{12.0}{\rmdefault}{\mddefault}{\updefault}{\color[rgb]{0,0,0}$a$}%
}}}}
\put(3736,-3031){\makebox(0,0)[lb]{\smash{{\SetFigFont{10}{12.0}{\familydefault}{\mddefault}{\updefault}{\color[rgb]{0,0,0}$s_1$}%
}}}}
\put(4501,-2941){\makebox(0,0)[lb]{\smash{{\SetFigFont{10}{12.0}{\familydefault}{\mddefault}{\updefault}{\color[rgb]{0,0,0}$b$}%
}}}}
\put(2521,-2311){\makebox(0,0)[lb]{\smash{{\SetFigFont{10}{12.0}{\familydefault}{\mddefault}{\updefault}{\color[rgb]{0,0,0}$\lnot a \land \lnot b$}%
}}}}
\put(2161,-2041){\makebox(0,0)[lb]{\smash{{\SetFigFont{10}{12.0}{\familydefault}{\mddefault}{\updefault}{\color[rgb]{0,0,0}$a$}%
}}}}
\put(2116,-3031){\makebox(0,0)[lb]{\smash{{\SetFigFont{10}{12.0}{\familydefault}{\mddefault}{\updefault}{\color[rgb]{0,0,0}$s_0$}%
}}}}
\put(3871,-3841){\makebox(0,0)[lb]{\smash{{\SetFigFont{10}{12.0}{\familydefault}{\mddefault}{\updefault}{\color[rgb]{0,0,0}$\lnot c$}%
}}}}
\end{picture}%

%% file: figures/dfa1.pdftex_t
\begin{picture}(0,0)%
\includegraphics{figures/dfa1.pdf}%
\end{picture}%
\setlength{\unitlength}{4144sp}%
\begingroup\makeatletter\ifx\SetFigFont\undefined%
\gdef\SetFigFont#1#2#3#4#5{%
  \reset@font\fontsize{#1}{#2pt}%
  \fontfamily{#3}\fontseries{#4}\fontshape{#5}%
  \selectfont}%
\fi\endgroup%
\begin{picture}(4973,4587)(-5819,-4975)
\put(-5804,-2176){\makebox(0,0)[lb]{\smash{{\SetFigFont{10}{12.0}{\familydefault}{\mddefault}{\updefault}{\color[rgb]{0,0,0}$a$}%
}}}}
\put(-4544,-556){\makebox(0,0)[lb]{\smash{{\SetFigFont{10}{12.0}{\familydefault}{\mddefault}{\updefault}{\color[rgb]{0,0,0}$a \land \lnot b$}%
}}}}
\put(-1574,-511){\makebox(0,0)[lb]{\smash{{\SetFigFont{10}{12.0}{\familydefault}{\mddefault}{\updefault}{\color[rgb]{0,0,0}$a \land b \land \lnot c$}%
}}}}
\put(-3329,-871){\makebox(0,0)[lb]{\smash{{\SetFigFont{10}{12.0}{\familydefault}{\mddefault}{\updefault}{\color[rgb]{0,0,0}$b \land c$}%
}}}}
\put(-5219,-1861){\makebox(0,0)[lb]{\smash{{\SetFigFont{10}{12.0}{\familydefault}{\mddefault}{\updefault}{\color[rgb]{0,0,0}$c \land \lnot b \land \lnot a$}%
}}}}
\put(-5579,-4246){\makebox(0,0)[lb]{\smash{{\SetFigFont{10}{12.0}{\rmdefault}{\mddefault}{\updefault}{\color[rgb]{0,0,0}$b \lor c$}%
}}}}
\put(-4904,-2986){\makebox(0,0)[lb]{\smash{{\SetFigFont{10}{12.0}{\familydefault}{\mddefault}{\updefault}{\color[rgb]{0,0,0}$\lnot a \land b$}%
}}}}
\put(-4184,-3436){\makebox(0,0)[lb]{\smash{{\SetFigFont{10}{12.0}{\familydefault}{\mddefault}{\updefault}{\color[rgb]{0,0,0}$a \land \lnot c$}%
}}}}
\put(-2834,-4111){\makebox(0,0)[lb]{\smash{{\SetFigFont{10}{12.0}{\familydefault}{\mddefault}{\updefault}{\color[rgb]{0,0,0}$c$}%
}}}}
\put(-3689,-4876){\makebox(0,0)[lb]{\smash{{\SetFigFont{10}{12.0}{\familydefault}{\mddefault}{\updefault}{\color[rgb]{0,0,0}$s_2$}%
}}}}
\put(-2744,-2536){\makebox(0,0)[lb]{\smash{{\SetFigFont{10}{12.0}{\familydefault}{\mddefault}{\updefault}{\color[rgb]{0,0,0}$\lnot a \land b \land \lnot c$}%
}}}}
\put(-1079,-1861){\makebox(0,0)[lb]{\smash{{\SetFigFont{10}{12.0}{\familydefault}{\mddefault}{\updefault}{\color[rgb]{0,0,0}$c$}%
}}}}
\put(-3329,-1456){\makebox(0,0)[lb]{\smash{{\SetFigFont{10}{12.0}{\familydefault}{\mddefault}{\updefault}{\color[rgb]{0,0,0}$a \land \lnot b \land \lnot c$}%
}}}}
\put(-5264,-3526){\makebox(0,0)[lb]{\smash{{\SetFigFont{10}{12.0}{\familydefault}{\mddefault}{\updefault}{\color[rgb]{0,0,0}$s_0$}%
}}}}
\put(-1394,-3121){\makebox(0,0)[lb]{\smash{{\SetFigFont{10}{12.0}{\familydefault}{\mddefault}{\updefault}{\color[rgb]{0,0,0}$s_4$}%
}}}}
\put(-1214,-1321){\makebox(0,0)[lb]{\smash{{\SetFigFont{10}{12.0}{\familydefault}{\mddefault}{\updefault}{\color[rgb]{0,0,0}$s_3$}%
}}}}
\put(-4364,-1321){\makebox(0,0)[lb]{\smash{{\SetFigFont{10}{12.0}{\familydefault}{\mddefault}{\updefault}{\color[rgb]{0,0,0}$s_1$}%
}}}}
\put(-5039,-4426){\makebox(0,0)[lb]{\smash{{\SetFigFont{10}{12.0}{\familydefault}{\mddefault}{\updefault}{\color[rgb]{0,0,0}$\lnot a \land b \land \lnot c$}%
}}}}
\end{picture}%

%% file: figures/ex1.pdftex_t
\begin{picture}(0,0)%
\includegraphics{figures/ex1.pdf}%
\end{picture}%
\setlength{\unitlength}{4144sp}%
\begingroup\makeatletter\ifx\SetFigFont\undefined%
\gdef\SetFigFont#1#2#3#4#5{%
  \reset@font\fontsize{#1}{#2pt}%
  \fontfamily{#3}\fontseries{#4}\fontshape{#5}%
  \selectfont}%
\fi\endgroup%
\begin{picture}(5415,1888)(-7,-926)
\put(316,-106){\makebox(0,0)[lb]{\smash{{\SetFigFont{10}{12.0}{\familydefault}{\mddefault}{\updefault}{\color[rgb]{0,0,0}$s_0$}%
}}}}
\put(1891,-106){\makebox(0,0)[lb]{\smash{{\SetFigFont{10}{12.0}{\rmdefault}{\mddefault}{\updefault}{\color[rgb]{0,0,0}$s_1$}%
}}}}
\put(1036,254){\makebox(0,0)[lb]{\smash{{\SetFigFont{10}{12.0}{\familydefault}{\mddefault}{\updefault}{\color[rgb]{0,0,0}$a \land r$}%
}}}}
\put(1801,839){\makebox(0,0)[lb]{\smash{{\SetFigFont{10}{12.0}{\rmdefault}{\mddefault}{\updefault}{\color[rgb]{0,0,0}$a \land \lnot b$}%
}}}}
\put(181,794){\makebox(0,0)[lb]{\smash{{\SetFigFont{10}{12.0}{\rmdefault}{\mddefault}{\updefault}{\color[rgb]{0,0,0}$\lnot r$}%
}}}}
\put(1126,-511){\makebox(0,0)[lb]{\smash{{\SetFigFont{10}{12.0}{\rmdefault}{\mddefault}{\updefault}{\color[rgb]{0,0,0}$\lnot a \land \lnot b$}%
}}}}
\put(1891,-871){\makebox(0,0)[lb]{\smash{{\SetFigFont{10}{12.0}{\rmdefault}{\mddefault}{\updefault}{\color[rgb]{0,0,0}$\lnot c$}%
}}}}
\put(3466,-106){\makebox(0,0)[lb]{\smash{{\SetFigFont{10}{12.0}{\rmdefault}{\mddefault}{\updefault}{\color[rgb]{0,0,0}$s_2$}%
}}}}
\put(4321,254){\makebox(0,0)[lb]{\smash{{\SetFigFont{10}{12.0}{\rmdefault}{\mddefault}{\updefault}{\color[rgb]{0,0,0}$c$}%
}}}}
\put(2656,254){\makebox(0,0)[lb]{\smash{{\SetFigFont{10}{12.0}{\rmdefault}{\mddefault}{\updefault}{\color[rgb]{0,0,0}$b$}%
}}}}
\put(5041,-106){\makebox(0,0)[lb]{\smash{{\SetFigFont{10}{12.0}{\rmdefault}{\mddefault}{\updefault}{\color[rgb]{0,0,0}$s_3$}%
}}}}
\end{picture}%

%% file: construction.tex
Given an SERE formula $\psi$, we want to efficiently construct a DFA $M$ 
with a number of states linear in the size of $\psi$ 
that is equisatisfiable to $\psi$ such
that the trace $\rho$ that satisfies $M$ also satisfies $\psi$.
We focus on the two sources of non-determinism:
the initial states and the $*$ operator. 

We first consider formulae $\psi$ of the form 
$\psi = x_1y_1; x_2y_2;\ldots;x_ny_n$ where $\ap$ and $\terms$ denote
the atomic propositions and SERE terms of $\psi$, respectively, 
$x_i\in \terms$, $y_i \in \set{\epsilon, *}$, and $\epsilon$ is the empty string.
We want to construct a DFA $M=(Q,I,F,\eval',\terms',\delta)$ 
where $Q=\set{s_0,s_1,\ldots,s_n}$, $I=\set{s_0}$,
and where each state $s_i$ corresponds to a term
$x_i$ in $\psi$.
The other components $F$, $\eval'$, $\terms'$, and $\delta$ will be discussed later. 

Consider the initial state $s_0$,
and consider an input valuation $v$ that matches $x_1$ the first term in $\psi$. 
The DFA $M$ needs to allow for two possibilities: (1) $v$ is part of the 
sequence matching the terms of $\psi$, 
and (2) $v$ is ignored and next inputs are considered as the match 
to the first term in $\psi$.
For example, consider $\psi=a;b$ where $\ap=\set{a,b}$ and consider the trace 
$\rho = \tpl{v_1,v_2,v_3,v_4}$ where
$v_1=\set{(a,\true),(b,\false)}$, $v_2=\set{(a,\true),(b,\false)}$,
$v_3=\set{(a,\true),(b,\true)}$, and $v_4=\set{(a,\false),(b,\true)}$.
The subtrace $\tpl{v_2,v_3}$ of $\rho$ matches $\psi$ while $v_1$ matches 
only the first term $a$ in $\psi$. 
Also $\tpl{v_3,v_4}$ matches $\psi$. 
The DFA $M$ should allow a choice of whether to stay in the initial state $s_0$
or to start the acceptance chain of transitions. 


Consider the subformula $x_1*;x_2$ which specifies that input valuations that 
match the term $x_1$ occur zero or more times in succession followed by a 
valuation that matches the term $x_2$. 
By definition, this includes non-determinism at every step.
Once the valuation that matches $x_1$ is presented, 
$M$ should allow for more valuations
matching $x_1$, and since we are restricting $s_1$ to correspond to $x_1$,
$M$ stays at the same state. 
$M$ should as well allow for valuations matching $x_2$
by transitioning to state $s_2$. 

Consider the SERE formula $\psi = a;b*;a$ where $\ap = \set{a, b,c}$. 
Consider the trace $\rho=\tpl{v_1,v_2,v_3,v_4,v_5}$ where
$v_1=\set{(a,\true),(b,\false)}$, $v_2=\set{(a,\false),(b,\true)}$,
$v_3=\set{(a,\true),(b,\true)}$,  $v_4=\set{(a,\false),(b,\true)}$, and
, $v_5=\set{(a,\true),(b,\false)}$. 
Again, $\rho$ can satisfy $\psi$ in several ways. 
One way is to consider accepting the subtrace $\tpl{v_1, v_2, v_3}$, 
and another is consider accepting the subtrace $\tpl{v_3, v_4, v_5}$. 
Once an input valuation such as $v_3$ that matches the second $a$ term 
in $\psi$ is presented, 
$M$ can move into the accepting state.
We use one free atomic proposition to allow the choices. 
It can also wait since $v_2\models b*$ and 
then upon receiving $v_5$ it will go to the accepting state. 

Further non-determinism needs to be considered when two consecutive 
terms in $\psi$ use the $*$ operator. 
For example, 
the input 
traces $\rho_1=\tpl{v_1,v_4}$ 
$\rho_2=\tpl{v_1,v_2,v_4}$,
$\rho_3=\tpl{v_1,v_3,v_4}$, and
$\rho_4=\tpl{v_1,v_2,v_3,v_4}$,
where
$v_1(a)=v_2(b)=v_3(c)=v_4(d)=\true$,
all satisfy the formula $\psi = a;b^{*};c^{*};d$. 
$M$ needs to allow for enough 
choices on the states corresponding to the $*$ terms to accept the 
four traces. 
For $m$ consecutive $*$ operators, $x_i*;x_{i+1}*;\ldots;x_{i+m-1}*$, 
we consider the corresponding $m$ states 
$S^*=\set{s_i, s_{i+1},\ldots, s_{i+m-1}}$ with all transitions possible from state 
$s_k\in S^*$ to state $s_p\in S^*$ where $k\le p$ on the same input valuation.
Therefore, we need $\lceil \log_2{m} \rceil$ atomic propositions to encode these
transitions as 
$(s_k,x_k\land \mathit{choice}(k,p,\bar{r}),s_p)$ where $\bar{r}$ is the 
vector of additional atomic propositions and $\mathit{choice}$ is a unique
choice of a valuation of propositions in $r$ mapped to $p$ and $k$. 
The same applies to terms in $\psi$ that follow $s_i$ 
such that $y_i = \epsilon$ and $y_{i-1} = *$.

%

\subsection{Equisatisfiable DFA construction}

Let $\ap'=\ap \cup \bar{r}$ where $\bar{r}$ is the vector of 
additional atomic propositions.
$\terms'$ is the set of SERE terms where $\ap'$ is the set 
of atomic propositions, and 
$\eval'$ is the set of valuations where $\ap'$ is the set of 
atomic propositions.
We construct the transition function $\delta$ by constructing 
four partial transition functions. 

The function $\delta_0$ denotes the transitions at the initial state.
\parbox{2.8in}{ \begin{tabbing}
mm\=mm\=mm\=mm\=mm \kill
$\delta_0$ \> $=$ \> $\set{ (s_0, \lnot r, s_0), (s_0, r \land x_1, s_1)}$.
\end{tabbing}
}

The function $\delta_{\epsilon}$ is the transitions corresponding to terms 
$x_iy_i$ $i\in [1\ldots n]$ where $y_i=\epsilon$ and $y_{i+1}\not=\epsilon$. 
{\small
\parbox{2.8in}{ \begin{tabbing}
mm\=mm\=mm\=mm\=mm \kill
$\delta_{\epsilon}$\>$=$ \>$~\{(s_i, x_{i+1}, s_{i+1})~|~ 0\le i < n$ and $s_i,s_{i+1} \in Q$ and \\
  \> \> \>$y_i = \epsilon \}$ \\
  \> $\union$ \>$\{(s_i, \lnot x_{i+1}, s_0,) ~|~  0\le i < n$ and $s_i, s_0 \in Q$ and  \\
  \> \> \> $y_i = \epsilon \}$
\end{tabbing}
}
}

The function $\delta_*$ is the transitions corresponding to terms
$x_iy_i$ $i\in [1\ldots n]$ where $y_i=*$. 
{\small
\parbox{2.8in}{ \begin{tabbing}
mm\=mm\=mm\=mm\=mm \kill
$\delta_{*}$ \>$=$ \> $\{ (s_i, \bigwedge_{j=i}^{n} \lnot x_j, s_0) ~|~ 
  (m \leq i \leq n)$ and $y_i = *$ and \\
  \> \> \> $y_m = \epsilon$ and $m \geq 1 \}$ \\
  \> $\union$ \> $\{ \big(s_i, x_j \land t_{ij}, s_j) ~|~ i\le j\le m \le n$ and \\
  \> \> \> $\forall k. i\le k < m \implies  y_k=*$ and \\
  \> \> \> $t_{ij} = \big(\mathit{choice}(i,j,\bar{r}) \lor \forall k. i\le k\le m\implies \lnot x_k$\big) \}
\end{tabbing}
}
}

The function $\delta_{\epsilon*}$ is the transitions corresponding to terms
$x_iy_i$ $i\in [1\ldots n]$ where $y_i=\epsilon$ and $y_{i+1}=*$. 
{\small
\parbox{2.8in}{ \begin{tabbing}
mm\=mm\=mm\=mm\=mm \kill
$\delta_{\epsilon*}$ \>$=$ \> $\{ (s_i, \bigwedge_{j=i}^{n} \lnot x_j, s_0) ~|~ 
  m\le i \le n$ and $y_i = *$ and \\
  \> \> \> $y_m = \epsilon$ and $m \geq 1 \}$ \\
  \> $\union$ \> $\{ \big(s_i, x_j \land t_{ij}, s_j) ~|~ i< j\le m \le n$ and \\
  \> \> \> $\forall k. i< k < m \implies y_k=*$ and \\
  \> \> \> $t_{ij} = \big(\mathit{choice}(i,j,\bar{r}) \lor \forall k. i< k\le m\implies \lnot x_k$ \big) \} \\
\end{tabbing}
}
}

The difference between $\delta_{\epsilon*}$ and $\delta_*$ is that in 
$\delta_{\epsilon*}$ no self transitions are defined. 

%
%

The transition function $\delta$ is now defined as
$\delta = \delta_0 \union \delta_\epsilon \union \delta_* 
\union \delta_{\epsilon*}$ .

Finally, we construct $F$ the set of accepting states as follows.
If $y_n = \epsilon$, then $F = \set{s_n}$.
If $y_n = *$ then 
$F=\{s_i ~|~ i=n $ or $k\le i\le n$ and $\forall j. k<j\le n\implies y_j=*\}$. 
Intuitively, this includes the states corresponding to the suffix of terms
with $*$ including one preceding term. 
For example, 
the accept states in $M$ corresponding to the formula 
$x_1;x_2;x_3*;x_4*$ 
are $F=\set{s_2,s_3,s_4}$.

\remove{
  Formally, given a SERE formula $\psi$ of the form $\psi = x_1y_1, x_2y_2, ..., x_ny_n$, we construct the DFA 
$M = \left(Q, I, F, \delta, \eval\right)$. We set $Q = \{s_0, s_1, ..., s_n\}$ to be the set of states of $M$. 
$I = \{s_0\}$ will be the set of initial states. We define the transition function $\delta$ as the union of three
transition functions $\delta_0, \delta_\epsilon, \delta_*$.  
Additionally, we define $R = \{ r_1, r_2, r_3, \ldots, r_p \}$ where $r_i$ is an auxiliary variable, and 
$p = \lceil \log_2{n+1-m} \rceil$ where $n$ is the number of states in $M_\psi$ and $m$ is 
the number of successive $*$ operators. 
We define $v$ to be a function $R \rightarrow \mathbb{B}$, and $[v]$ to represent all 
of the bit representations of each $r_i \in R$. Therefore $|[v]| = 2 ^ {\log_2{n+1-m}} = n + 1 -m$.
The three transition functions $\delta_0, \delta_\epsilon$ and $\delta_*$ are shown in 
equation~\ref{eq:construct1}.
}

\begin{theorem}[Equisatifiability of $M$ and $\psi$]
\label{thm:equisat}
A formula $\psi=x_1y_1;x_2y_2;\ldots;x_ny_n$ 
where $\ap=\set{x_1,x_2,\ldots x_n}$, 
and a constructed DFA $M=(Q,I,F,\eval',\terms',\delta)$,
$M$ and $\psi$ are equisatisfiable. 
In addition if there exists $\rho$ that satisfies  $M$ then 
$\rho$ also  satisfies $\psi$.
\end{theorem}

The proof is by induction on the length of the formula $\psi$ and 
is available in the online appendix$^{\ref{fn:url}}$.
Note that it is shown in the proof that the satisfiability of 
$\psi$ and $M$ will be by the same trace, with some existential 
quantification over the added free (auxiliary) atomic propositions.

\subsection{Input to ABC }


The ABC solver accepts an And-Inverted-Graph (AIG) sequential circuit as input. 
An AIG is a sequential circuit restricted to only use 
AND and NOT logical gates. 
The translation from a DFA to an equivalent AIG circuit is straightforward.
In short, we encode each state from the DFA by a unique valuation of the 
register variables, 
and construct the initial values and the next state functions of the
registers according to $\delta$. The AIG circuit will also have a
unique output $o$ who is true only when the values of the registers correspond
to a state in $F$. Note that $o$ will be then negated in order to 
perform bounded model checking. 


For a formula of the form $\psi=\phi_1 \land \phi_2$,
we construct $C_1$ and $C_2$ that correspond to $\phi_1$ and 
$\phi_2$, respectively, and we use a conjunction of the 
outputs of $C_1$ and $C_2$ to correspond to the satisfiability 
of $\psi$.
Similarly, we use a disjunction for $\phi_1 \lor \phi_2$.

%% file: implementation.tex
We implemented our technique and integrated it with the 
ABC~\cite{brayton2010abc} synthesis and verification framework.
We used ANTLR~\cite{parr1995antlr} to provide users with a C like
input language, augmented with constructs that support
wire declarations, synchronization,  and SERE specifications. 
Our tool supports scalar variables, 
boolean variables, arrays, and functions including recursion.

The tool generates an AIG circuit as discussed in Section~\ref{s:core}. 
The added free atomic propositions are left as free 
primary input variables into the AIG circuit. 


The goal of the verification procedure is to ensure that there exists at least
one setting of the primary input variables that leads the AIG representing 
the SERE specification $\psi$
from its initial state to one of its accept states. 
Let $\mathbb{R}$ be the 
set of all possible valuations of $\bar{r}$;
$|R|\le m$ where $m$ is the maximum number of 
consecutive $*$ operators in $\psi$ 
since the size of $\bar{r}$ is bounded by $\log_2(m)$.
Our goal is to prove
that $\exists{v_r \in \mathbb{R}}$ such that $\psi$ is satisfied.
We encode the existential quantifier with a disjunction over all 
the valuations in $\mathbb{R}$. 


If the system under test violates $\psi$, ABC returns a counterexample and our 
tool provides a user friendly debugging interface to debug the system.
Before performing symbolic or bounded model checking, 
the user can make use of the ABC framework to perform circuit level 
optimizations, an advantage 
not present in traditional model checking tools such as 
NuSMV\cite{cimatti2002nusmv}. 
This can help reduce the size of the problem. 
For bounded model checking, the user can also provide a bound
on the number of transitions of the system. 
ABC will then check that the specification $\psi$ is always valid 
within the provided upper bound. 

\begin{figure}[bt]
\begin{tabular}{p{3.7cm} p{3.7cm}}
\begin{Verbatim} [fontsize=\relsize{-2.5}]
int x;
x = 0;
while ( true ) {
 @do_together {
  if ( x == 3 )
   x = 0;
  else 
   x = x + 1;
@guarantee_sere_invariant
    cntr; } }
\end{Verbatim} 
&
\begin{Verbatim} [fontsize=\relsize{-2.5}]
@sere cntr {
 atoms x0, x1, x2, x3, x4.
 x0 <- (x == 0).
 x1 <- (x == 1).
 x2 <- (x == 2).
 x3 <- (x == 3).

 Formula f.
 f = (x0;x1;x2;x3;x0).
}
\end{Verbatim}
\end{tabular}
\caption{Example of a 2 bit counter}
\label{fig:example}
\end{figure}

Figure~\ref{fig:example} shows the implementation of a $2$ bit counter.
The \cci{@do\_together} modifier denotes that
the enclosed list of statements occur simultaneously. 
The \cci{@guarantee\_sere\_invariant} is a synchronization constructs 
that times the specification evaluation.
The \cci{@sere} block lists the specifications.
Atoms $x_i, 0 \leq i \leq 3$  evaluate to \cci{true} when $x=i$. 

%% file: results.tex
\input{results_table}

We compare our implementation with NuSMV2~\cite{cimatti2002nusmv}, a symbolic 
model checker
used for the verification of system designs. 
NuSMV2 accepts Computational Tree Logic,
Property Specification Language, and LTL as specification languages. 
We compare our implementation with the NuSMV2 model checker for LTL properties. 

In several examples, such as the load balancer example, 
we succeeded to generate an AIG and find counterexamples in defect circuits
where other techniques in NuSMV2 failed.

All computation times provided in the following are obtained on a machine 
with 2.20 Ghz Intel Core i7 processors running an x64-version of Ubuntu Linux.
The allowed memory usage is up to 8 GB and we set a timeout of 1800 seconds. 
For our experiments, we used NuSMV v2.5.4. 

\subsection{LILY~\cite{jobstmann2006optimizations} and UNBEAST~\cite{ehlers2010symbolic} Examples}
We used LILY examples~\cite{jobstmann2006optimizations} 
and the UNBEAST load balancer example%
~\cite{ehlers2010symbolic} as benchmarks for comparison. 
We passed LTL formulae from the benchmarks to LILY and UNBEAST and 
generated implementation designs. 
Then we translated the resulting designs manually into the input language of our 
tool as well as into SMV, the language of NuSMV2.
In the cases where LILY and UNBEAST were not able to generate designs, 
we manually wrote dummy designs in which defects surely exist. 
Note that in both cases, we manually translated the LTL properties 
into SERE.

We passed the implementation annotated with the SERE specification to 
both NuSMV and to our tool and compared the results based on the size 
of the resulting structure passed to the model checker, and on 
the computation time. 
NuSMV generates BDDs to perform reachability analysis.
Our tool generates AIG circuits and passes them to ABC. 
We used the number of latches and AND gates in our synthesized AIG 
before and after applying optimizations versus the number of states 
in the generated DFA and the total number of BDD nodes from NuSMV2.
We use the commands \cci{dump\_fsm} and
\cci{print\_usage} to obtain such information from the NuSMV2 tool.

Table~\ref{t:results} shows a summary of the results obtained from performing 
formal verification of the realizable load balancing examples 
from UNBEAST~\cite{ehlers2010symbolic} and 
examples from the LILY suites~\cite{jobstmann2006optimizations}. 
Designs labeled 
as \texttt{load\_*} correspond to load balancing examples, while designs labeled 
as \texttt{demo-v*} correspond to examples from the LILY benchmarks corresponding 
to a traffic light system. 
Note that we restrict our attention to realizable LTL formulae.

We used the demos, version 3 and 19, from the LILY benchmarks for comparison. 
We were able to generate the circuits and verify them in both cases. 
We employed several circuit level synthesis
techniques available ABC~\cite{brayton2010abc} and were able to 
significantly reduce the size of the problem. 
NuSMV2 was also able to verify both models efficiently.

For the load balancer examples, 
we verified 5 out of the examples that we tested and we found
problems and fixed them in the others. 
We used LILY and UNBEAST to generate the models from the specifications, 
and then checked the generated
models against their specifications. 
NuSMV2 was also able to verify the 5 examples
but failed (timed out at 30 minutes) to synthesize the LTL formulae for other 
benchmarks such as 
\cci{load\_30, load\_75}, \cci{load\_76},
\cci{load\_76}, \cci{load\_77}, \cci{load\_78},
and \cci{load\_79}.
UNBEAST and LILY were not
able to generate a model of the specifications as well. 
Notice that the  \cci{load\_79} benchmark is the largest design
in the load balancer benchmarks with $9$ clients and a fixed 
number of servers.

This is evidence of the high utility of our technique
which enables model checking where other tools fail. 
We also note that in all of the cases, the size of the problem we 
send to the model checker was smaller than the size of the 
problem generated by NuSMV2.

\subsection{IBM Arbiter case Study}
\input{ibm_table}

We also used our tool against the IBM generalized buffer~\cite{IBM_arbiter}. 
The model consists of four senders
that communicate with a generalized buffer in order to send data to two receivers. 
Each sender has its own
data line while the receivers share a common data bus. 
The buffer also includes a first-in first-out queue. 
We translated the VHDL implementation
provided from IBM and checked it against the defined specifications. 
We checked for two assertions on the design. 
\begin{itemize}
  \item Sender requests are always acknowledged, and 
  \item arbiter requests are always acknowledged.
\end{itemize}

Note that since the original LTL assertions are of the 
form {\em ``is always acknowledged''}, writing an SERE specification
for the good traces would not be useful for bounded model checking since 
the specification would match if one request was acknowledged once. In order
to overcome this limitation, we can use a bound on the number of requests and
then check that all requests within this bound have been acknowledged. 

We were able to efficiently verify the first assertion. 
However when verifying the second assertion 
our implementation detected a counter example, and after 
debugging and inspection we found that there is a defect in 
the assignment of the request acknowledgments in the provided
VHDL implementation. 
Table~\ref{t:ibm} shows the size of the synthesized AIG 
circuit in terms of number of latches and number of 
And gates before and after optimizations, and the verification decision 
of our tool for both specifications. 

Our tool and the experiments are all available online$^{\ref{fn:url}}$.

%% file: results_table.tex
\begin{table*}[tb]\scriptsize
\begin{center}
\begin{tabular}{|c|c|c|c|c|c|c|c|c|c|c|}
\cline{2-11}
\multicolumn{1}{l|}{} & \multicolumn{6}{c|}{{\bf Our tool}} & \multicolumn{4}{c|}{{\bf NuSMV}} \\ \cline{2-11}
\multicolumn{1}{l|}{} & \multicolumn{2}{c|}{{\bf Synthesis}} & \multicolumn{2}{c|}{{\bf Optimizations}} & \multirow{2}{*}{{\bf Verification}}%
& \multirow{2}{*}{\bf Total Time(s)} & \multicolumn{2}{c|}{{\bf Synthesis}} & \multirow{2}{*}{{\bf Verification}} & \multirow{2}{*}{{\bf Total Time(s)}} \\ \cline{1-5} \cline{8-9}
{\bf Design}  & {\bf latches} & {\bf Ands} & {\bf latches} & {\bf Ands} & & & {\bf States} & {\bf BDD nodes} &  &  \\ \hline
Load\_0 & 29 & 289 & 0 & 0 & Verified & 0 & 10 & 453 & Verified & 0.004 \\ \hline
Load\_7 & 87 & 1018 & 20 & 72 & Verified & 0 & 32 & 13171 & Verified & 0.04 \\ \hline
Load\_8 & 27 & 271 & 0 & 0 & Verified & 0 & 10 & 387 & Verified & 0.004 \\ \hline
Load\_24 & 63 & 902 & 26 & 131 & Verified & 0.07 & 2481 & 48053 & Verified & 1.228 \\ \hline
Load\_30 & 193 & 3089 & 123 & 584 & Found counter & 0.05 & \multicolumn{2}{c|}{Timeout}& \multicolumn{2}{c|}{NA}   \\ \hline
Load\_75 & 110 & 1476 & 49 & 203 & Found counter & 0.18 & \multicolumn{2}{c|}{Timeout} & \multicolumn{2}{c|}{NA}  \\ \hline
Load\_76 & 124 & 1832 & 64 & 301 & Found counter & 0.19 & \multicolumn{2}{c|}{Timeout} & \multicolumn{2}{c|}{NA}  \\ \hline
Load\_77 & 137 & 2046 & 69 & 321 & Found counter & 0.32 & \multicolumn{2}{c|}{Timeout} & \multicolumn{2}{c|}{NA}  \\ \hline
Load\_78 & 151 & 2270 & 75 & 359 & Found counter & 0.36 & \multicolumn{2}{c|}{Timeout} & \multicolumn{2}{c|}{NA}  \\ \hline
Load\_79 & 164 & 2488 & 80 & 381 & Found counter & 0.38 & \multicolumn{2}{c|}{Timeout} & \multicolumn{2}{c|}{NA}  \\ \hline
demo-v3 & 48 & 639 & 31 & 151 & Verified & 0.11 & 70 & 4237 & Verified & 0.012 \\ \hline
demo-v19 & 66 & 941 & 48 & 634 & Verified & 0.08 & 96 & 9332 & Verified & 0.008 \\ \hline
\end{tabular}
\end{center}
\caption{{\scriptsize Results of our tool compared to NuSMV}}
\label{t:results}
\end{table*}

%% file: ibm_table.tex
\begin{table*}[tb] \scriptsize
\begin{center}
\begin{tabular}{|c|c|c|c|c|c|c|}
\cline{2-7}
\multicolumn{1}{l|}{} & \multicolumn{2}{c|}{{\bf Synthesis}} & \multicolumn{2}{c|}{{\bf Optimizations}}%
& \multicolumn{2}{c|}{{\bf Verification}} \\ \hline
{\bf Assertion} & {\bf Latches} & {\bf Ands} & {\bf Latches} & {\bf Ands} & {\bf Verification} & {\bf Time(s)} \\ \hline
(1) & 800 & 3209 & 0 & 0 & Verified & 0 \\ \hline
(2) & 792 & 3139 & 38 & 141 & Found counter & 0.61 \\ \hline
\end{tabular}
\end{center}
\caption{{\scriptsize Size of the problem and decision of our tool on the IBM GenBuf Arbiter}}
\label{t:ibm}
\end{table*}

%% file: related.tex

Several techniques have been developed in the literature in order to synthesize LTL formulae, 
usually describing properties that hold over real-life hardware systems and designs. These synthesis
techniques have different targets, some aim to generate complete and correct systems based on 
input specifications, while others are targeted at generating monitors to ensure correct
functionality of systems through assertion checking.
We differ than most of the literature in that we synthesize an equisatisfiable circuit
to the formula that is good to be used for model checking purposes only. 

NuSMV2~\cite{cimatti2002nusmv} is a symbolic model checking tool that employs both 
satisfiability (SAT) and BDD based model checking techniques. It processes an input 
describing the logical system design as a finite state machine, and a set of specifications
expressed in LTL, Computational Tree Logic and Property Specification Language.
Given a model $M$ and a set of specifications $P$, NuSMV2 first flattens $M$ and $P$ by 
resolving all module instantiations and creating modules and processes, thus generating one 
synchronous design. It then performs a boolean encoding step to eliminate all scalar variables, 
arithmetic and set operations and thus encode them as boolean functions.   

In order to avoid the state space explosion problem, NuSMV2 performs a cone of 
influence reduction~\cite{berezin1998compositional} step in order to eliminate
non-needed parts of the flattened model and specifications. The cone of influence
reduction abstraction technique aims at simplifying the model in hand by only 
referring to variables that are of interest to the verification procedure, i.e. variables
that influence the specifications to check~\cite{clarke1999model}. We use
NuSMV2 to compare the results of our implementation on a set of benchmarks as
described in Section~\ref{s:results}.

FoCs is an industrial tool developed at IBM research labs, targeted at generating simulation 
checkers from formal specifications~\cite{p:Focs}. The tool's goal is to reduce, or possibly eliminate the 
amount of human intervention in writing and maintaining functional checkers. FoCs takes input specification
expressed in RCTL~\cite{p:rctl}, and generates formal checkers written in VHDL. These checkers
are then linked with the original VHDL and executed on a set of test programs. The role of the formal checkers 
is to make sure that the original design never goes into an error state. 

The generation of the formal checkers from the RCTL specifications is done in three steps. First, the RCTL is 
translated into a NDFA according to the algorithm described in~\cite{p:rctl}. This NDFA will have a set of error
states, which represent the states that the design should never go into if it meets the required specifications.  
In order to be able to generate the VHDL checkers, the NDFA has to be translated into a DFA, which is in turn translated
into a VDHL process. 
This process will then be run alongside the original design to check for 
violations of the specifications. 

The key drawback of FoCs' approach is that transformation algorithm generates a DFA that can be exponential 
in the number of states of the NDFA, which takes us back to the state-space explosion problem. The authors claim 
that such a limitation does not exist in their case, since the simulation is rather sensitive to the number 
of VHDL lines in the generated checker, which is at most quadratic in the size of the property to check. 
Our approach differs from FoCs in that it aims at generating a AIG 
free primary input variables that is linear in 
the size of the property, without generating an intermediary NDFA. 
Therefore, it can help rendering the generated VHDL checker 
even smaller in terms of the lines of code. 

Jobstmann et. al developed LILY~\cite{jobstmann2006optimizations}, a synthesis tool aimed at synthesizing correct
designs from LTL specifications. It is implemented on top of Wring, and introduces several 
optimizations based on alternating tree automata, covering both game based
and simulation based optimization techniques. They present an incremental algorithm for
checking realizability of LTL formulae, and output a Verilog~\cite{thomas2002verilog} 
model in case the formula is realizable. We made use of LILY to generate several design models, 
and then we checked these generated models against their original specifications using our
own implementation.  

UNBEAST~\cite{ehlers2010symbolic} is a synthesis tool that aims to generate system designs that 
are correct by construction. It takes as input a specification containing environment assumptions 
and system guarantees, and splits them into safety and non-safety conditions. Each of these
sets of conditions are then handled differently in the synthesis game. Unlike LILY, it relies
on universal co-B\"{u}chi word automata instead of co-B\"{u}chi tree automata. It checks for 
realizability of LTL formulae and returns SMV models when realizable. 
We differ from both UNBEAST and LILY in the type and the goal of synthesis. Our goal is to generate
monitor from SERE properties, while LILY and UNBEAST generate models that satisfy the LTL properties. 
Our generated DFA is equisatisfiable to the input SERE property, and thus can be used for model 
checking purposes only. 

%% file: conclusion.tex
In this paper we presented a technique 
that takes a formula in SERE and transforms it into an AIG
circuit with a number of states that is linear in terms 
of the length of the formula. 
The generated circuit is equisatisfiable to the formula
and enables the use of symbolic model checking and 
bounded model checking where it was not possible before;
i.e. where the typical translation from NDFA equivalents of the formula
to a DFA blows up exponentially.